\RequirePackage{fix-cm}

\documentclass{svjour3}                     
    
\usepackage{graphicx}
\usepackage{bm}
\usepackage{amsmath}
\usepackage{amsfonts}
\usepackage{siunitx}
\usepackage{xcolor}
\journalname{myjournal}

\begin{document}

\title{Reconstructing velocity and pressure from sparse noisy particle tracks using Physics-Informed Neural Networks}
\titlerunning{Reconstructing velocity and pressure from data}

\author{Patricio Clark Di Leoni$^{*,1}$          \and
        Karuna Agarwal$^{2}$ \and
        Tamer A.~Zaki$^{2}$\and
        Charles Meneveau$^{2}$\and
        Joseph Katz$^{2}$
}

\institute{$^{*}$ pclarkdileoni@udesa.edu.ar\\
$^{1}$ Departamento de Ingeniería,
Universidad de San Andrés, Victoria, Buenos Aires, Argentina \\\
$^{2}$ Department of Mechanical Engineering, Johns Hopkins University,
              Baltimore, MD 21218, USA}

\date{Received: date / Accepted: date}

\maketitle

\begin{abstract}

Volume-resolving imaging techniques are rapidly advancing progress in experimental fluid mechanics.
However, reconstructing the full and structured Eulerian velocity and pressure fields from  sparse and noisy particle tracks obtained experimentally remains a significant challenge.  
We introduce a new method for this reconstruction, based on Physics-Informed Neural Networks (PINNs). 
The method uses a Neural Network regularized by the Navier-Stokes equations to interpolate the velocity data and simultaneously determine the pressure field. 
We compare this approach to the state-of-the-art Constrained Cost Minimization method \cite{agarwal_reconstructing_2021}.
Using data from direct numerical simulations and various types of synthetically generated particle tracks, we show that PINNs are able to accurately reconstruct both velocity and pressure even in regions with low particle density and small accelerations. 
PINNs are also robust against increasing the distance between particles and the noise in the measurements, when
studied under synthetic and experimental conditions.

\end{abstract}

\section{Introduction}

Unstructured particle trajectory information measured in a fluid flow experiment through particle tracking techniques \cite{dabiri_particle_2019}, such as the Shake-the-Box method \cite{schanz_shake--box_2016}, can be used to calculate velocity statistics and Lyapunov exponents, find Lagrangian coherent structures \cite{haller_lagrangian_2015}, or estimate helicity \cite{angriman_broken_2021}, for example. More challenging, however, is the calculation of spatial gradients or reconstruction of the pressure field. Procedures to infer the pressure usually involve
first interpolating the velocity field and then solving the pressure-Poisson equation \cite{ghaemi_piv-based_2012,kat_pressure_2012,oudheusden_piv-based_2013,villegas_evaluation_2014}. Other methods rely on calculating the material acceleration, either through a pseudo-Lagrangian 
\cite{jensen_optimization_2004,liu_instantaneous_2006,liu_vortex-corner_2013} or Eulerian approach
\cite{violato_lagrangian_2011} from the interpolated velocity fields, or directly from the tracks 
\cite{novara_particle-tracking_2013,schanz_shake--box_2016}, and then integrating the pressure gradients \cite{baur_piv_1999,dabiri_algorithm_2014} with, for example, an omni-directional method
\cite{liu_instantaneous_2006,wang_gpu-based_2019}.

The problem of reconstructing flow fields from partial measurements can be approached using a variety of data assimilation techniques \cite{zaki_limited_2021}. 
Ensemble- and adjoint-variational methods \cite{mons_kriging-enhanced_2019,buchta_2021,Buchta2022,wang_spatial_2019,Wang_adjoint_2019} and nudging approaches \cite{clark_di_leoni_inferring_2018,clark_di_leoni_synchronization_2020,wang_synchronization}, can mix Eulerian and Lagrangian data, and work under turbulent conditions, but usually require numerically solving sets of partial differential equations on a grid.  On the other hand, new machine learning methods, such as convolutional neural networks \cite{fukami_super-resolution_2019,callaham_robust_2019}, generative adversarial networks \cite{xie_tempogan_2018,buzzicotti_reconstruction_2020,oh_accurate_2022} and deep operator networks \cite{cai_deepmmnet_2021,mao_deepmmnet_2020,clark_di_leoni_deeponet}, have been applied to the reconstruction of two-dimensional images and low-dimensional flow fields.

In this paper we focus on Physics-Informed Neural Networks \cite{raissi_physics-informed_2019}, a class of artificial neural networks designed to approximate physical fields and whose training is informed by physics of the problem. The physics-information comes from adding the residuals of the equations of motion of the problem, calculated directly from the network using automatic differentiation \cite{goodfellow_deep_2016}, to the loss function as a regularization term and therefore the governing equations are weakly enforced.  The network architecture can be designed to enforce hard constraints, for example predictions of solenoidal fields \cite{du2021ednn}.  Two notable advantages are that PINNs do not require data on a regular grid and do not require information on the material acceleration. PINNs have been shown to be effective in inverse problems in fluid and solid mechanics \cite{raissi_hidden_2018,shukla_physics-informed_2020} and have been used to reconstruct velocity and pressure fields in Tomographic Background Oriented Schilieren measurements \cite{cai_flow_2021}.  A detailed comparison of PINN and adjoint-variational data assimilation (4DVar) in turbulent channel flow was performed in \cite{du_VarPINN2022}.  

We apply the PINN technique to both synthetic and experimental datasets and compare to the state-of-the-art Constrained Cost Minimization (CCM) technique \cite{agarwal_reconstructing_2021}. The synthetic datasets are from Direct Numerical Simulation of a turbulent channel flow, where the particle trajectories are either obtained directly from the simulation or via a synthetic tomography procedure \cite{schanz_shake--box_2016} to mimic error sources in experimental measurements. The experimental dataset consists of velocity measurements in the shear layer that develops behind a backward-facing step. We compare errors and correlations in the streamwise velocity field, the pressure and their respective gradients, as well as calculate the temporally resolved spectra, and provide quantitative comparisons with CCM.

\section{Physics-Informed Neural Networks}

\begin{figure}[h]
    \centering
    \includegraphics[width=0.95\textwidth]{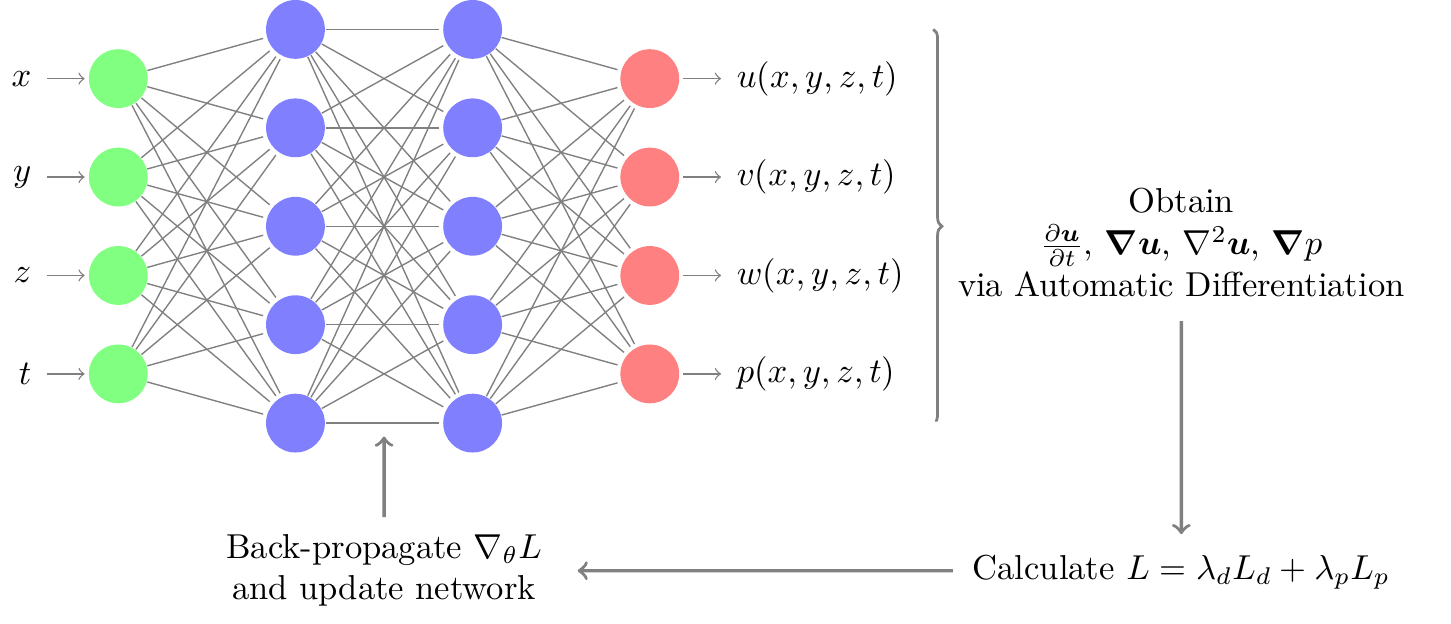}
    \caption{Diagram of the PINN and its training procedure}
    \label{fig:diag}
\end{figure}

Physics-Informed Neural Networks \cite{weinan2018deep,raissi_hidden_2018,raissi_physics-informed_2019} are designed to describe a set of physical fields and make use of the partial differential equations that govern these fields to constrain and regularize the training process.  
In the present application the PINNs enforce the three-dimensional, incompressible Navier-Stokes equations, with the coordinates $(x,y,z,t)$ as inputs and $(u,v,w,p)$ as the outputs evaluated at the given coordinate.  The architecture is a typical fully-connected neural network with parameters $\bm{\theta}$ and where every hidden unit is passed through an activation function $\sigma$, as shown in the diagram in Fig.~\ref{fig:diag}(a).   

The goal is to take a set of velocity-field measurements $\hat{\Omega}_d = \{x_j, y_j, z_j, t_j; \hat{u}_j, \hat{v}_j, \hat{w}_j\}^{N_d}_{i=1}$, and use the PINN to interpolate these data. The \textit{Physics-Informed} property is achieved by applying automatic differentiation to calculate the derivatives of the outputs of the network with respect to its inputs and then evaluating the terms in the Navier-Stokes equations; the residual of the equations is included in the loss function of the network as a regularization term. The loss function is thus composed of two terms: The first compares the network predictions to the measurement data, 
\begin{equation}
    L_d = \frac{1}{N_d} \sum^{N_d}_{j=1} \left\vert \bm{u}_j - \hat{\bm{u}}_j \right\vert^2 ,
\end{equation}
where $\Omega_d = \{x_j, y_j, z_j, t_j; u_j, v_j, w_j\}^{N_d}_{i=1}$ are the input-output pairs of the PINN and where the coordinates points $(x_j, y_j, z_j, t_j)$ in $\Omega_d$ and $\hat{\Omega}_d$ coincide. The second loss function is associated with the physics, and is comprised of two contributions, one from the residual of the incompressibility condition,
\begin{equation}
    L_i = \frac{1}{N_p} \sum^{N_p}_{j=1} \left\vert
              \bm{\nabla}\cdot \bm{u}_j \right\vert^2,
\end{equation}
and the other from the residual of the momentum equations,
\begin{equation}
    L_m = \frac{1}{N_p} \sum^{N_p}_{j=1}  \left\vert
              \frac{\partial \bm{u}_j}{\partial t} +
              \bm{u}_j\cdot \bm{\nabla} \bm{u}_j +
              \frac{1}{\rho} \bm{\nabla} p_j -
              \nu \nabla^2 \bm{u}_j
              \right\vert^2.
              \label{eq:Lmerror}
\end{equation}
Note that the set $\Omega_p = \{x_j, y_j, z_j, t_j; u_j, v_j, w_j, p_j\}^{N_p}_{j=1}$ does not necessarily have to coincide or overlap with $\Omega_d$. Finally, the total loss function takes the form,
\begin{equation}
    L = 
    \underbrace{\lambda_d L_d}_{\text{data part}} +
    \underbrace{\lambda_i L_i + \lambda_m L_m}_{\text{physics part}},
\end{equation}
where $\lambda_d$, $\lambda_i$ and $\lambda_m$ are, in general, independent hyper-parameters used to balance each term of the loss function. Since the loss can be arbitrarily normalized, one of these hyper-parameters can be set to unity. For simplicity, we refer to the physics part as $L_p = \lambda_i L_i + \lambda_m L_m$. Further details on how to choose these hyper-parameters are given below.

It is important to remark that all the derivatives in $L_p$ are
calculated through automatic differentiation on the network. Automatic
differentiation is the process by which the derivative of a function
composed of combinations and concatenations of known elementary functions
(the activation functions for the case of neural networks)
is calculated using the chain rule. It is the same process used to
calculate the gradients of the loss function with respect to the weights 
of the network when training a normal neural network, but in
PINNs it is also applied to the inputs of the network. 

\subsection{Balancing the terms of the loss function}

Including a regularization term in the loss function of a neural network
introduces additional weights, or hyper-parameters, to be prescribed.
Recent works have presented different strategies: some based on the
analysis of the Hessian of the loss function
\cite{wang_understanding_2020}, others based on Neural Tangent Kernel
Theory \cite{wang_when_2020}, and others that incorporate the weighting
hyperparameters into the trainable parameters
\cite{mcclenny_self-adaptive_2020}. In this work we use the first of
these methods, but aided by our knowledge of the physical problem and its governing equations.  The
flows we present here have a mean component whose magnitude is on the order of 
$U$ along direction $\hat{x}$, so if we recognize the streamwise
advection term to be the leading term in the momentum equation, we obtain
the following scalings for each term in the loss function, 
\begin{gather}
   L_d \sim U^2,
   \\
   L_i \sim \left( \frac{\partial u}{\partial x} \right)^2,
   \\
   L_m \sim U^2 \left( \frac{\partial u}{\partial x} \right)^2.
\end{gather}
In highly anisotropic flows it is reasonable to separate the data and momentum terms into the three different components. While we do not separate them since in Eq.~\ref{eq:Lmerror} the norm includes all components, we do apply input and output normalization layers separately for each velocity component (introduced in the following section) to alleviate the effects of anisotropy.

Since in all the flows that we present $U\approx1$, $L_i$ and $L_m$ are of the same order, we therefore take $\lambda_i = \lambda_m = 1$  and balance the loss function by varying only $\lambda_d$, which according to our analysis should be of the order, 
\begin{equation}
\lambda_d \sim \frac{L_p}{L_d}  
   \sim \left( \frac{\partial u}{\partial x} \right)^2,
   \label{eq:ld_est}
\end{equation}
in order for every term of the loss function to be balanced.
This gradient can be estimated very approximately using Kolmogorov's turbulence. At scale $\ell$ the gradient magnitude is on the order
\begin{equation}
   \frac{\partial u}{\partial x} \sim \epsilon^{1/3} \ell^{-2/3},
\end{equation}
where $\epsilon$ is the rate of energy dissipation. This estimate takes its largest value at the smallest scale in the flow, i.e., the Kolmogorov scale $\eta \sim (\nu^3/\epsilon)^{1/4}$. Thus, we obtain

\begin{equation}
    \frac{\partial u}{\partial x} \sim \epsilon^{1/2} \nu^{-1/2}.
   \label{eq:grad_est}
\end{equation}

Since the convergence of the training procedure is ultimately dictated by the respective gradients of different terms in the loss function
\cite{wang_understanding_2020,wang_when_2020}, we cannot set the values
of the weighting hyper-parameters solely based on the above scaling alone.  For this reason, we factorize $\lambda_d$ into  $\lambda_d=\lambda_d^0 \hat{\lambda}_d$, where $\lambda_d^0$ is a fixed part whose value is inspired by Eqs.~\eqref{eq:ld_est} and \eqref{eq:grad_est}, and $\hat{\lambda}_d$ is a varying part set by the algorithm presented in \cite{wang_understanding_2020}, which updates $\hat{\lambda}_d$ at epoch $n$ according to, 
\begin{equation}
    \hat{\lambda}^n_d = (1-\alpha) \hat{\lambda}^{n-1}_d + \alpha
    \frac{\langle \vert \bm{\nabla}_\theta L_p \vert \rangle_\theta}{\langle
    \vert \lambda_d^0 \bm{\nabla}_\theta L_d \vert \rangle_\theta},
\end{equation}
where the superscripts $n$ and $n-1$ denote the value of $\hat{\lambda}_d$ at the $n$th and $(n-1)$th iteration, $\alpha$ is a new free hyperparameter usually set to $0.1$, and $\langle \vert \bm{\nabla}_\theta \cdot \vert \rangle_\theta$ is the mean of the absolute value of the gradients of each quantity with respect to the network parameters $\theta$. 

\begin{figure}[h]
    \centering
    \includegraphics[width=0.95\textwidth]{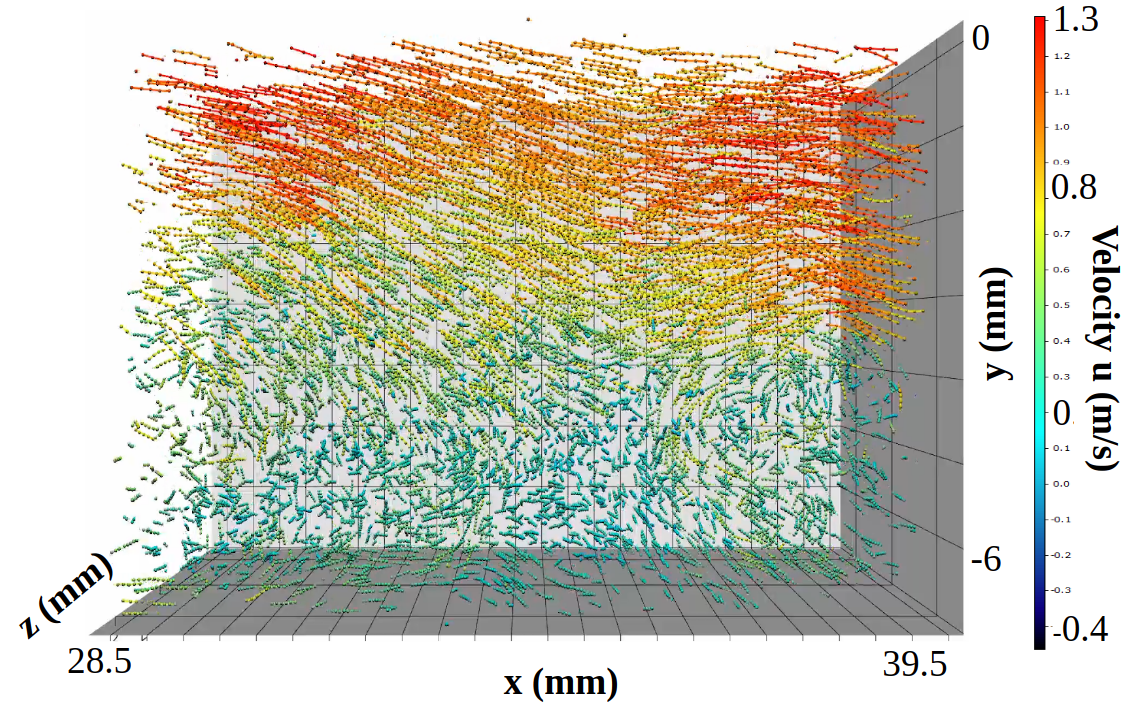}
    \caption{Particle tracks obtained from Case 3.}
    \label{fig:tracks}
\end{figure}

\subsection{Normalization layers}

Due to the nature of activation functions and in order to mitigate the differences in the orders of magnitudes of various inputs and also outputs, we added an input and an output normalization layer. Both layers aim to ensure that the values for each input or output channel are within the range $(-1,1)$ and are nearly symmetrized.

For the input part we use min-max normalization, which takes the minimum and maximum value for each coordinate and rescales the coordinate between $-1$ and $1$, namely (for coordinate $x$ for example)
\begin{equation}
    x \leftarrow 2 \frac{x - x_{\min}}{x_{\max} - x_{\min}} - 1,
\end{equation}
where $x_{\min}$ and $x_{\max}$ are the minimum and maximum values of $x$ in our domain, respectively. As the flow domain is known, all the respective maximums and minimums are also known.

For the output part we use z-score normalization, which centers and standardizes each output, for example
\begin{equation}
    u \leftarrow \sigma_u u + \mu_u,
\end{equation}
where $\mu_u$ and $\sigma_u$ are the mean and standard deviation of $u$,
respectively, and the $u$ on the RHS is the output of the last trainable
layer. As we use velocity measurements to train the networks, the values of
$\mu_u$, $\mu_v$, $\mu_w$, $\sigma_u$, $\sigma_v$, and $\sigma_w$ can be
easily estimated. Since no data is assumed to be known regarding pressure, the values of $\mu_p$ and $\sigma_p$ are left as free hyperparameters of the network.

\section{Methods and dataset description}

We analyze three different cases: the first two use data from a direct
numerical simulation (DNS) of a turbulent channel flow at $Re_\tau =
1000$ available through the Johns Hopkins Turbulence Database (JHTDB) \cite{Lietal2008,graham2016web,ChannelFlow1000},
and the third case studies experimental measurements of a turbulent
shear layer \cite{agarwal_reconstructing_2021}. While the sources and acquisition methods differ between cases, in all tests the data are particle tracks, as shown in Fig.~\ref{fig:tracks}. We define $x, y, z$ as the streamwise, vertical and spanwise directions, respectively. All volumes expressed below are in $\{x,y,z\}$ order. Below we discuss each dataset and detail the PINN hyperparameters used for each case.

\subsection{Case 1: Synthetic particle tracks in DNS of turbulent channel flow} 

The goal of the first case is to study the effects of particle spacing
and noise level on the accuracy of flow reconstruction. All quantities are
given in terms of the friction velocity $u_{\tau}$ and the viscous length
scale $\delta_\nu=\nu/u_\tau$ where $\nu$ is the fluid's kinematic viscosity. 
To generate the dataset we randomly seeded a volume
of size $235 \delta_\nu \times215 \delta_\nu \times185 \delta_\nu$, in the streamwise, vertical and spanwise directions, respectively,
approximately $10^{-5}$ of the total DNS volume, located at the bottom
wall of the channel.  The interrogation volume has dimensions
$195\times195\times145$ wall units and is slightly smaller than the
seeded one so as to avoid edge effects. The DNS has a constant horizontal resolution of
$12.3\delta_\nu$ in the streamwise direction and $6.1\delta_\nu$ in the
spanwise direction, while its vertical resolution varies along the height of the channel, it is close to $0.02\delta_\nu$ at the bottom of the test volume and close to $3.7\delta_\nu$ at the top.  The results are available every 5 frames of the
original DNS times steps. The JHTDB uses fourth order Lagrange
polynomials in space and a piecewise-cubic interpolation scheme in time
to interpolate data from grid locations to the spatio-temporal position of interest. The seeded particles are treated as Lagrangian tracers and the synthetic tracks are thus generated by tracking the particles' evolution in the volume. 
Several sets of tracks were generated, each composed of nine
snapshots (exposures), centered around a target time and with timestep
$dt=0.00325$ (or $2.5$ times the DNS timestep). The velocity and
acceleration of each particle were then calculated by fitting a second
order polynomial around the target time. All nine snapshots were used for the reconstruction using PINNs, while only three snapshots (the target time plus and minus one time step) were used when using CCM.
Noise was introduced to the measurements by adding a random Gaussian fluctuation to each particle position along the tracks.  
The flow reconstruction was performed in a total 39 different and independent target times at various values of particle spacing and noise levels. The particle spacing was controlled by varying the number of particles in the flow and it
ranged from $3.4\delta_\nu$ (when using $60{,}000$ particles) to $8.4\delta_\nu$ (when using $4{,}000$ particles). The noise levels ranged from zero (no noise) to $0.8$ pixels (equivalent to $0.32\delta_\nu$).

The PINNs used to reconstruct this case were eight layers deep and 200 units wide. Their initial learning rate was set to $10^{-3}$ and then followed an exponential decay schedule with rate $0.9$ and $100$ epochs characteristic time. The pressure was scaled with $\sigma_p=0.1$ and $\lambda_d^0$ was set to unity. The sets of points used to enforce the data and physics parts of the loss function, $\Omega_d$ and $\Omega_p$, respectively, coincided. The networks were trained for 750 epochs.

\subsection{Case 2: Synthetic tomographic images from DNS of channel flow}

In this case we use the same flow configuration and fields as in the previous one, i.e., the turbulent channel flow data from the JHTDB. 
We generate synthetic tomographic images of the particle fields using the EUROPIV Synthetic Image Generator \cite{lecordier_europiv_2004} and apply the Shake-the-box algorithm to reconstruct the tracks \cite{schanz_shake--box_2016} instead of using the particle positions directly. This procedure includes all the imaging, calibration and reconstruction errors encountered experimentally. 
The synthetic tomography is generated by projecting the particle positions onto four views, with all views aligned in the wall-normal direction and forming angles of $\pm {15}^\circ$ and $\pm {30}^\circ$ in the spanwise direction. 
The images are then processed with LaVision's DaVis 10 software to retrieve particle positions. 
Further details can be found in \cite{agarwal_reconstructing_2021}. 
Contrary to the previous case, we now reconstruct an extensive flow history consisting of 2{,}000 timesteps with $dt=0.00325$. 
The sample volume has dimensions $95\delta_\nu \times 95 \delta_\nu \times 45 \delta_\nu$ and the mean particle spacing is $5\delta_\nu$, which is maintained constant by re-feeding particles into the flow over a test volume larger than the sample volume.  
The velocity (and accelerations for the CCM) were calculated by fitting a second order
polynomial on 17 exposures.

Due to the size of the dataset, two PINNs were used to reconstruct the full $2{,}000$ frame long time window, each tasked with $1{,}000$ non-overlapping frames. As before, the PINNs used were eight layers deep and 200 units wide. Their initial learning rate was set to $10^{-3}$ and then followed an exponential decay schedule with rate $0.9$ and $100$ epochs characteristic time. The pressure was scaled with $\sigma_p=0.1$ and $\lambda_d^0$ was set to unity. The set of points used to enforce the physics, $\Omega_p$ contained all of the points in $\Omega_d$ as well as $N_d$ additional points randomly selected throughout the spatio-temporal domain. The networks were trained for 1{,}000 epochs.

\subsection{Case 3: Experimental data measured in a turbulent shear layer} 

\begin{figure}[t]
    \includegraphics[width=0.9\textwidth]{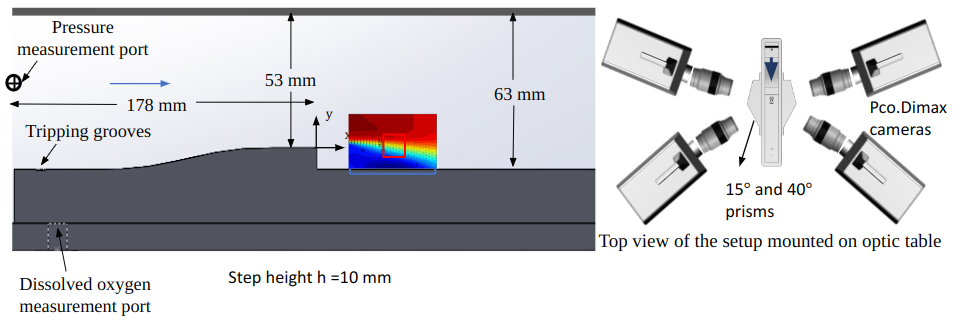}
    \caption{Experimental setup used for Case 3. The area of interest is marked with the red square. The colored square shows a 2D PIV visualization of the streamwise velocity.}
    \label{experimental_setup}
\end{figure}

For the third case we use experimental data from a flow behind a step in a small water tunnel. Figure~\ref{experimental_setup} shows the configuration. The  Reynolds number of the boundary layer upstream of separation is $Re_\tau=800$ and the flow presents strong vortical structures.

The test section is ${405 \times 63 \times 51}~\si{mm^3}$ and the step height $h$ is equal to $\SI{10}{mm}$. Four Pco.dimax cameras were used to record the images of size $624\times380~\si{pixel}$ at $\SI{14925}{Hz}$ over a field of view of ${12.5 \times 7.5 \times 4.5}~{mm^3}$ located in a region behind the step.  The free-stream velocity is $\SI{5.3}{m/s}$, which leads to high acquisition frequencies and low spatial resolution of images. As a  result relatively sparse particle spacing of $\sim\SI{300}{\micro\metre}$ are obtained.  The time window chosen for reconstruction is $900$ frames long.

Similar to the synthetic camera settings adopted in Case 2, the four cameras were at $\pm {15}^\circ$ and $\pm {40}^\circ$ angles in the spanwise direction. A Photonics DM60-527  Nd:YLF laser was used to illuminate the flow field.  The particles were $\SI{13}{\mu m}$ silver-coated hollow glass spheres. As in Case 2, the tomographic PTV data were processed with the Shake-the-Box algorithm from DaVis 10.  For more details on the experimental setup see \cite{gopalan_flow_2000}, and for more details on these particular measurements see \cite{agarwal_reconstructing_2021}.

Due to the size of the dataset, nine PINNs were used to reconstruct the full $900$ frame long time window, each tasked with $100$ non-overlapping frames. The PINNs used were six layers deep and 350 units wide. Their initial learning rate was set to $10^{-3}$ and then followed an exponential decay schedule with rate $0.9$ and $50$ epochs characteristic time. The pressure was scaled with $\sigma_p=1$ and $\lambda_d^0$ was set to $10^7$. The set of points used to enforce the physics, $\Omega_p$, contained all of the points in $\Omega_d$ plus $N_d$ additional points randomly selected throughout the whole spatiotemporal domain. The networks were trained for 2{,}000 epochs.

\begin{figure}[t]
    \centering
    \includegraphics[width=0.32\textwidth]{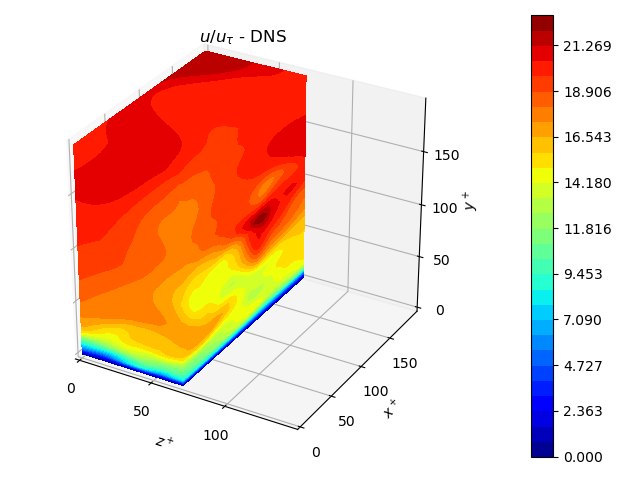}
    \includegraphics[width=0.32\textwidth]{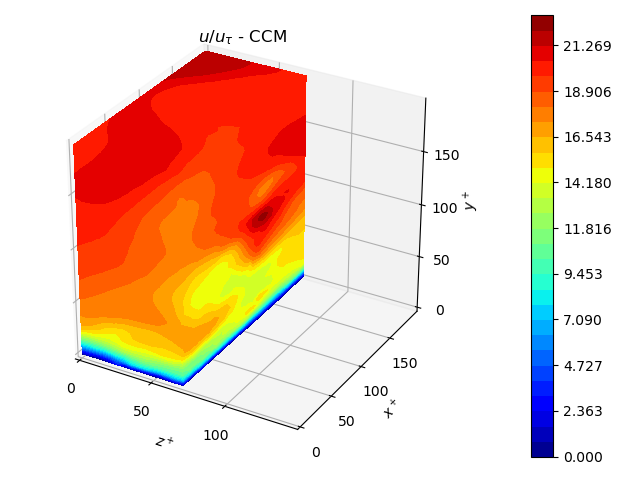}
    \includegraphics[width=0.32\textwidth]{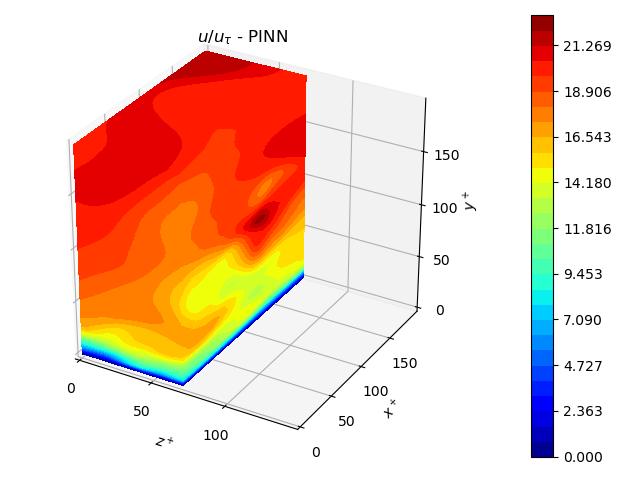}
    \\
    \includegraphics[width=0.32\textwidth]{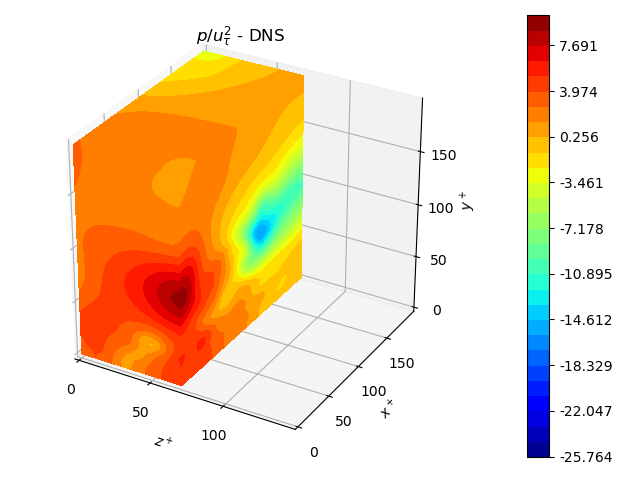}
    \includegraphics[width=0.32\textwidth]{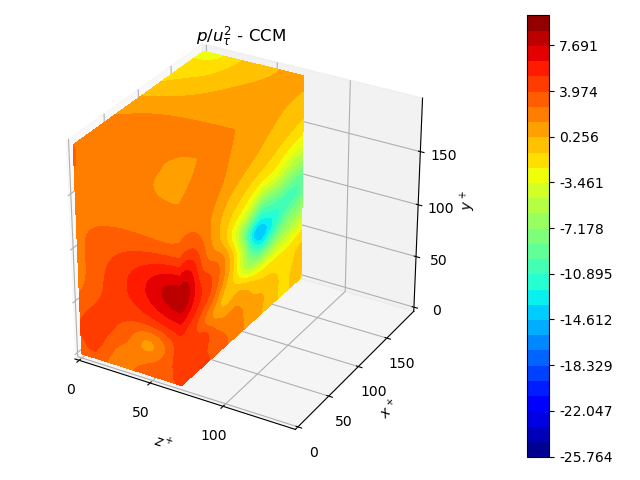}
    \includegraphics[width=0.32\textwidth]{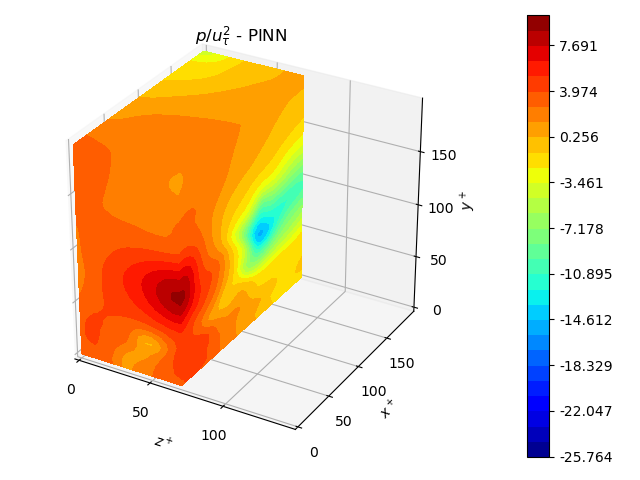}
    \caption{Instantaneous visualization of the data and the CCM and PINN
    reconstruction for Case 1. Top row: streamwise velocity field $u$. Bottom row:
    pressure $p$. Only half of the reconstructed volume is shown.}
    \label{viz_case1}
\end{figure}

\section{Results}

We now present results from the three test cases. To present quantitative comparisons between DNS and measured values for Cases 1 and 2 (where we have the ``truth'' from the DNS data), we use the root mean square error (RMSE) and correlation coefficients. The root-mean-square error has the following definition:

\begin{equation}
    \epsilon_u = \frac{\sqrt{\langle (u - u^{DNS})^2 \rangle}}{u_\tau} , ~~~~~~~~~~~~~~
    \epsilon_p = \frac{\sqrt{\langle (p - p^{DNS})^2 \rangle}}{u^2_\tau} ,
\end{equation}
where the averaging operation $\langle\cdot\rangle$ is performed over the entire spatiotemporal domain. In cases where the averaging operation is not performed over a particular dimension, this will be stated explicitly, for example $\epsilon_u (y^+)$ is the vertical profile of the RMSE of $u$ which is not averaged in the vertical dimension.
The correlation coefficients between DNS and measured data are defined by
\begin{equation}
    \rho_u = \frac{
    \langle
    (u - \langle u \rangle)
    (u^{DNS} - \langle u^{DNS} \rangle)
    \rangle}{
    \langle
    (u - \langle u \rangle)^2
    \rangle
    \langle
    (u^{DNS} - \langle u^{DNS} \rangle)^2
    \rangle
    } ,
\end{equation}
and $\rho_p$ follows the same definition as $\rho_u$.

\subsection{Case 1: Synthetic particle tracks in DNS of turbulent channel flow}

\begin{figure}[h]
    \centering
    \includegraphics[width=0.6\textwidth]{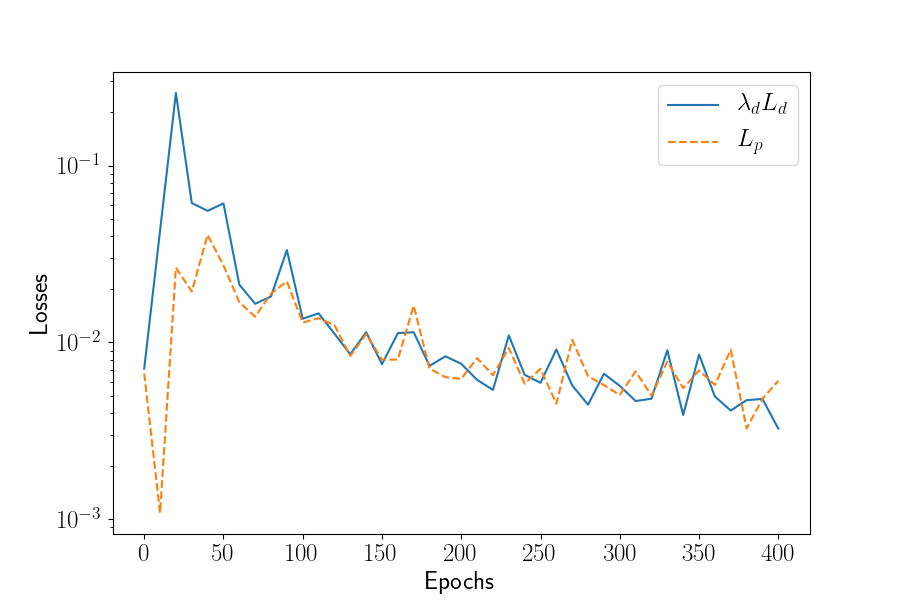}
    \caption{Evolution of the data and physics losses during the training of the PINN that produced the results of Fig.~\ref{viz_case1}.}
    \label{losses}
\end{figure}

We start by considering the results from Case 1.
In Fig.~\ref{viz_case1} we show visualizations of the streamwise velocity field $u$ and the pressure $p$ of the true data, the CCM-reconstructed field and the PINN-reconstructed field at one particular instant. 
The visualizations show only half the volume in order to demonstrate the quality of the reconstruction within the bulk of the volume. 
Both CCM and PINNs are in good qualitative agreement with the true data, with PINNs producing slightly smoother fields than CCM. In Fig.~\ref{losses} we show the evolution of the weighted data loss $\lambda_d L_d$ and the physics loss $L_p$ during the training of the PINN. Both losses are minimized by the training procedure and the balancing term $\lambda_d$ helps keep both terms of the same order.

\begin{figure}[t]
    \includegraphics[width=0.32\textwidth]{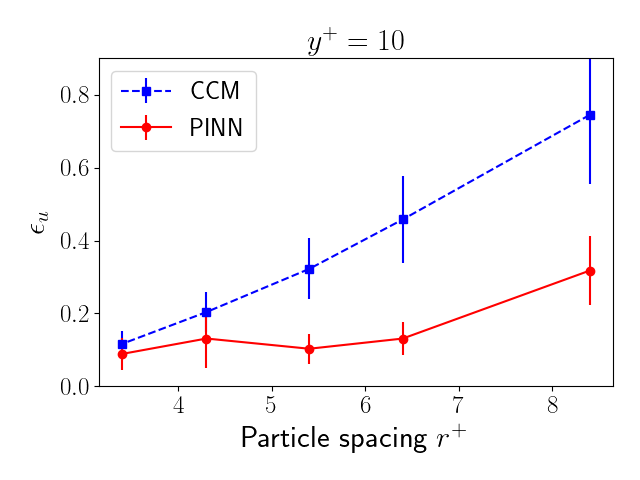}
    \includegraphics[width=0.32\textwidth]{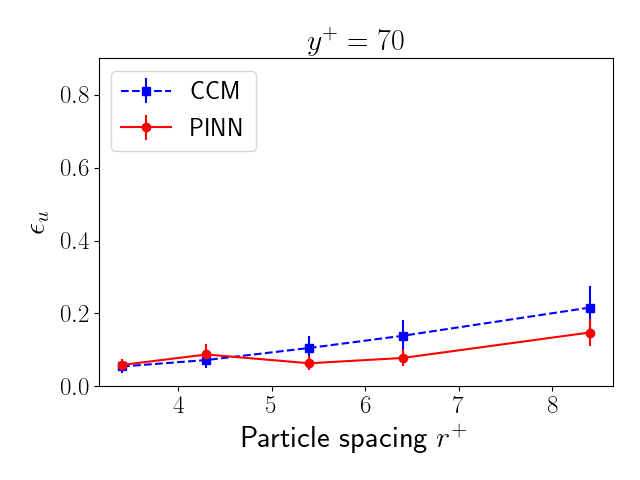}
    \includegraphics[width=0.32\textwidth]{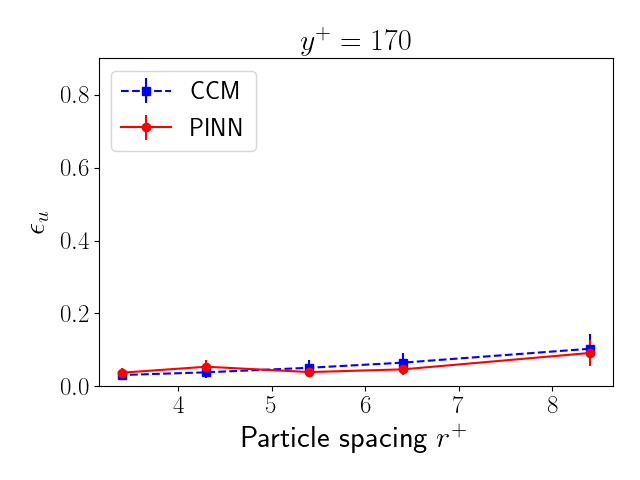}
    \caption{RMSE $\epsilon_u$ in Case 1 as a function of particle spacing
    without any added noise at three different heights.}
    \label{rmse_u_er_0}
\end{figure}

\begin{figure}[t]
    \includegraphics[width=0.32\textwidth]{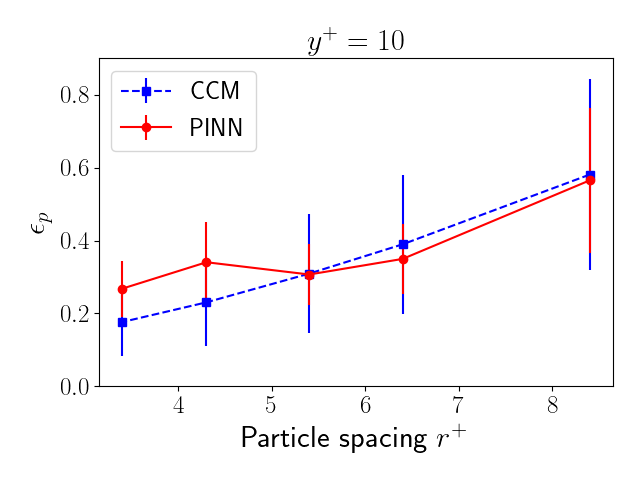}
    \includegraphics[width=0.32\textwidth]{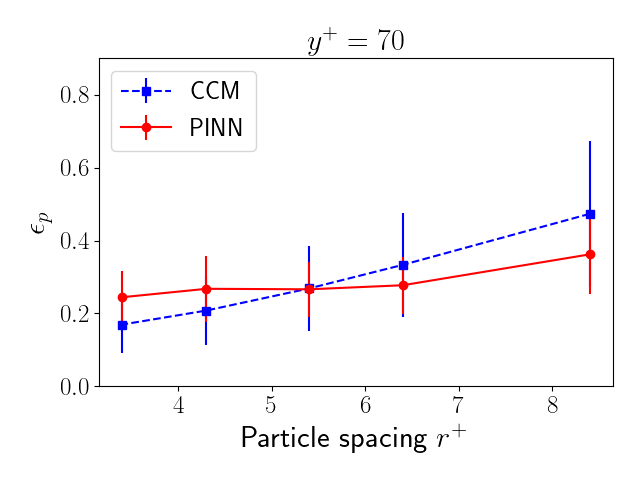}
    \includegraphics[width=0.32\textwidth]{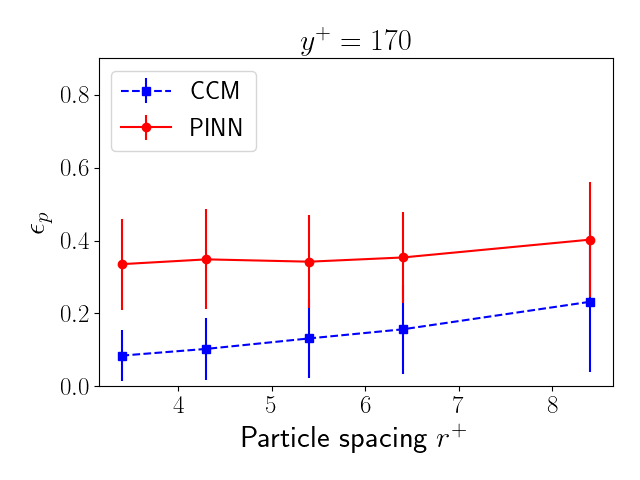}
    \caption{RMSE $\epsilon_p$ in Case 1 as a function of particle spacing
    without any added noise at three different heights.}
    \label{rmse_p_er_0}
\end{figure}

For a quantitative assessment, we examine the accuracy of the reconstruction as a function of the particle spacing. 
In Fig.~\ref{rmse_u_er_0} we show  $\epsilon_u$ as function of the particle spacing in absence of any measurement noise, at three different heights, and in Figure~\ref{rmse_p_er_0} we show the same for $\epsilon_p$. 
Results are averaged over the 39 independently reconstructed snapshots, and all error bars are equal to the calculated standard deviations. 
Both CCM and PINNs techniques are have commensurate success in reconstructing the true flow, with PINNs being slightly more robust to particle spacing for reconstructing $u$ and CCM yielding slightly smaller errors for $p$ near the top boundary (the omni-directional integration used by CCM is well-suited for boundaries since it iteratively solves for boundary pressures to match with interior material acceleration information). 

\begin{figure}[t]
    \includegraphics[width=0.32\textwidth]{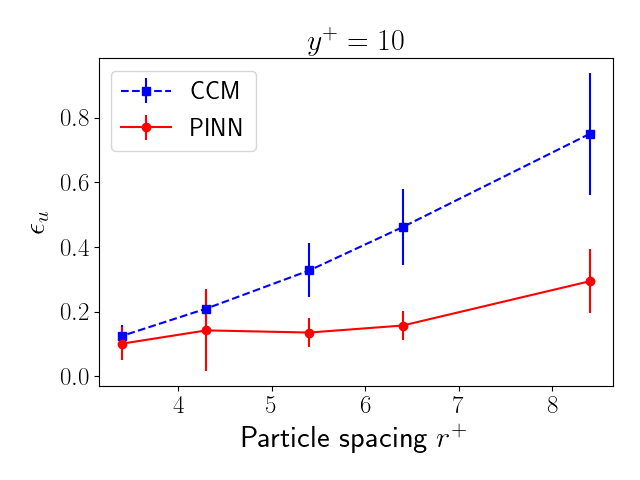}
    \includegraphics[width=0.32\textwidth]{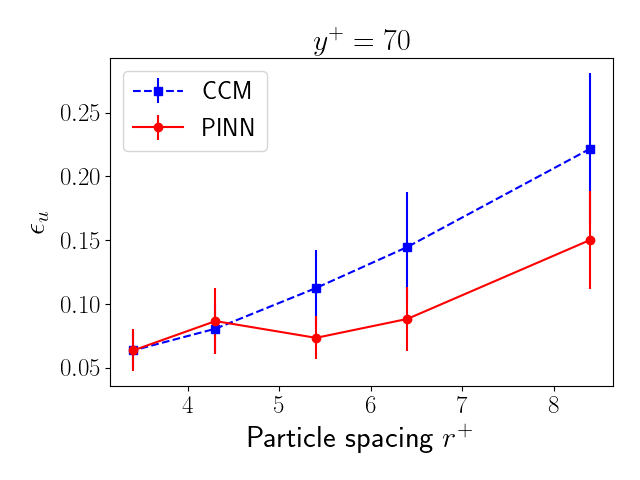}
    \includegraphics[width=0.32\textwidth]{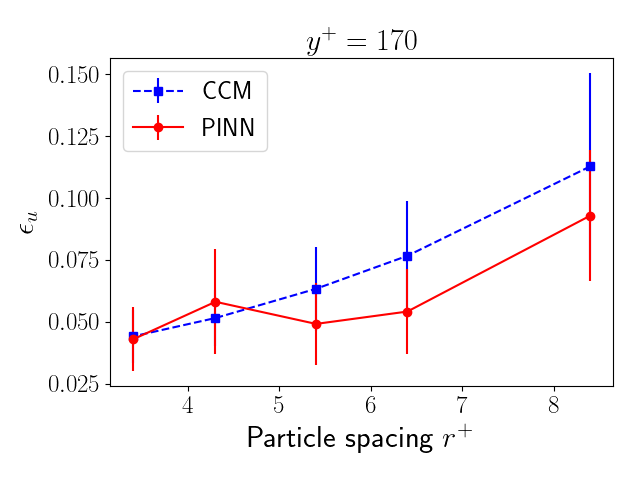}
    \caption{RMSE $\epsilon_u$ in Case 1 as a function of particle spacing
    with $\SI{0.4}{px}$ noise added at three different heights.}
    \label{rmse_u_er_4}
\end{figure}

\begin{figure}[t]
    \includegraphics[width=0.32\textwidth]{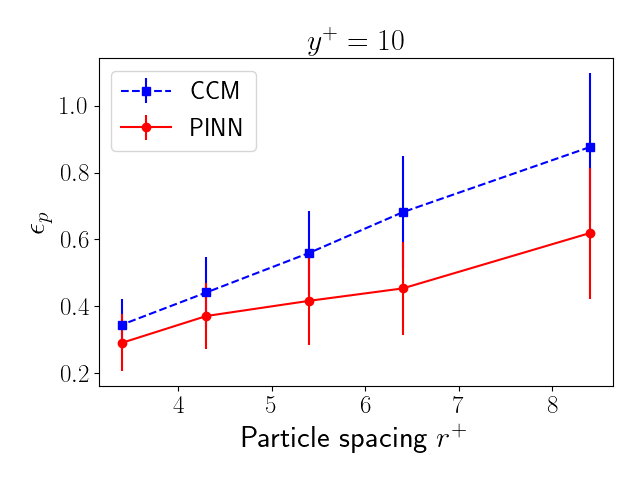}
    \includegraphics[width=0.32\textwidth]{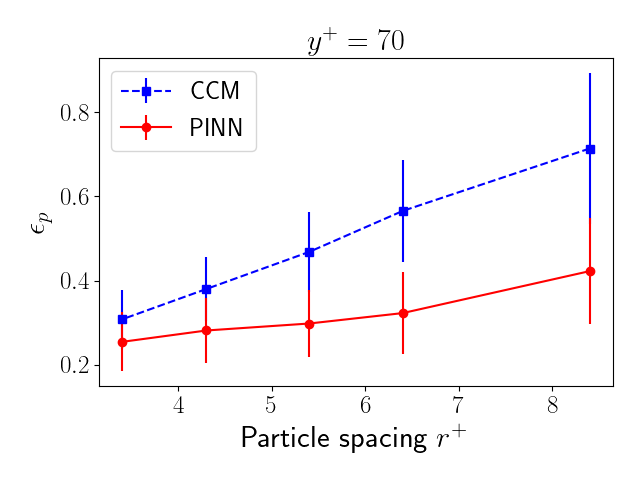}
    \includegraphics[width=0.32\textwidth]{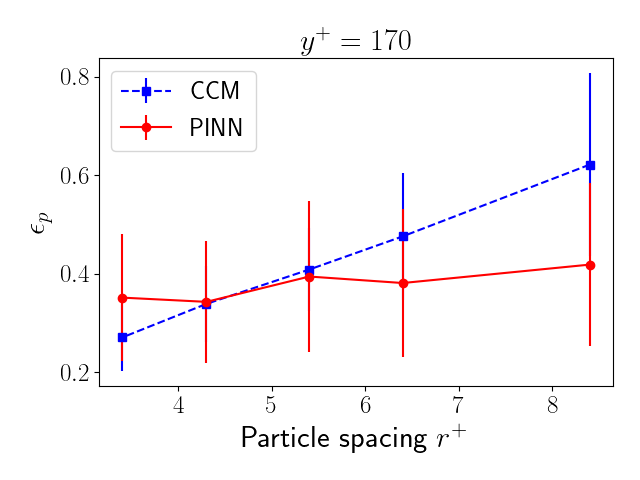}
    \caption{RMSE $\epsilon_p$ in Case 1 as a function of particle spacing
    with $\SI{0.4}{px}$ noise added at three different heights.}
    \label{rmse_p_er_4}
\end{figure}

\begin{figure}[t]
    \includegraphics[width=0.32\textwidth]{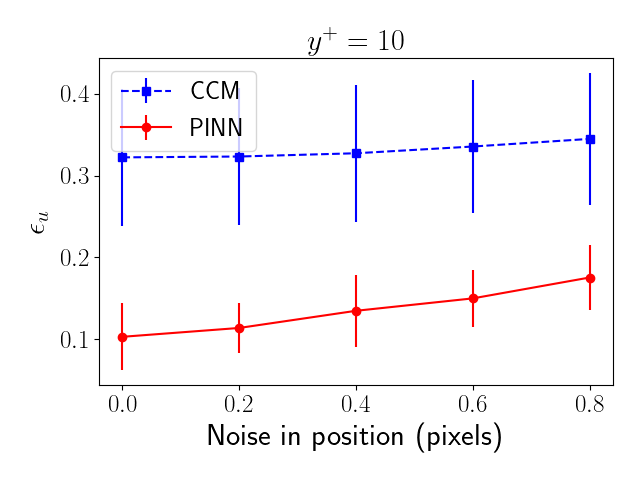}
    \includegraphics[width=0.32\textwidth]{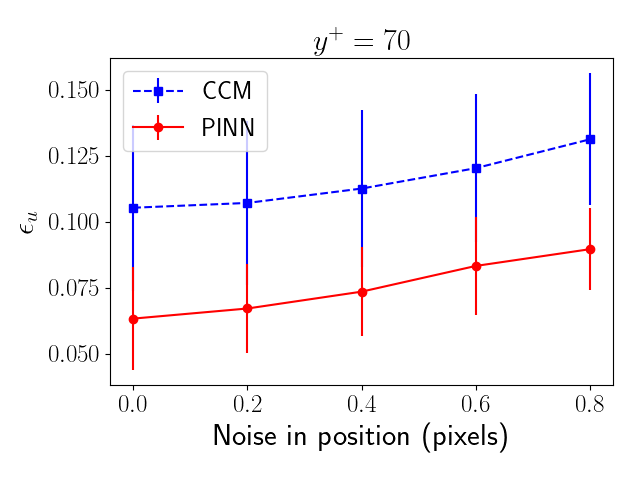}
    \includegraphics[width=0.32\textwidth]{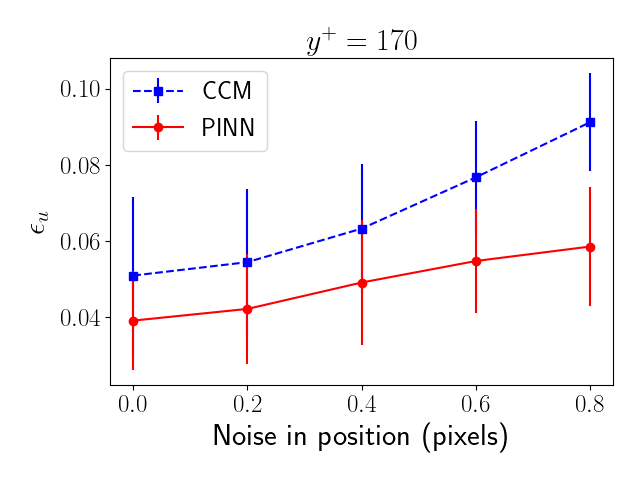}
    \caption{RMSE $\epsilon_u$ in Case 1 as a function of noise added
    with a particle spacing of $\SI{5.4}{\delta_\nu}$ at three different heights.}
    \label{rmse_u_np_15}
\end{figure}

\begin{figure}[t]
    \centering
    \includegraphics[width=0.32\textwidth]{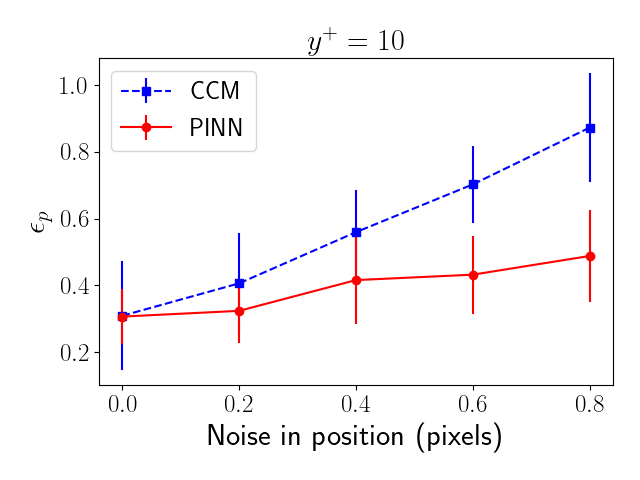}
    \includegraphics[width=0.32\textwidth]{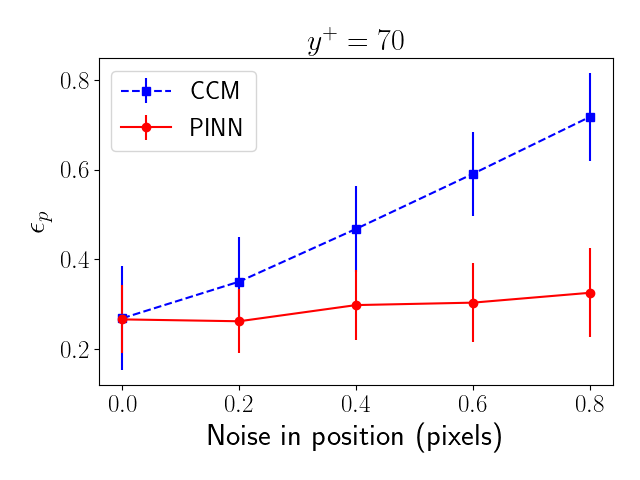}
    \includegraphics[width=0.32\textwidth]{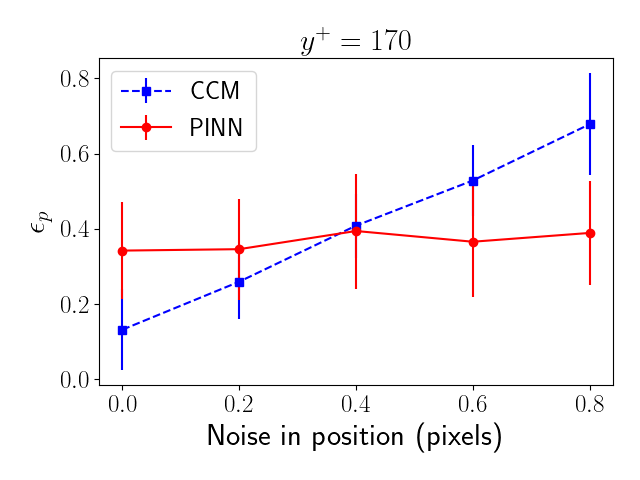}
    \caption{RMSE $\epsilon_p$ in Case 1 as a function of noise added
    with a particle spacing of $\SI{5.4}{\delta_\nu}$ at three different heights.}
    \label{rmse_p_np_15}
\end{figure}

In Figures~\ref{rmse_u_er_4} and \ref{rmse_p_er_4} we again show $\epsilon_u$ and $\epsilon_p$, respectively, at different locations, but this time with an added noise of $0.4$ pixels (equivalent to $0.16\delta_\nu$). Right away we can appreciate how PINNs are more robust than CCMs in noisy systems. To take this point further, in Figures~\ref{rmse_u_np_15} and \ref{rmse_p_np_15} we show $\epsilon_u$ and $\epsilon_p$, respectively, at three different heights but this time at a different noise levels, all with constant particle spacing $r^+=5.4$. The errors in the reconstruction generated with PINNs increase more slowly with noise level than in the case of CCM, especially for pressure. This is because the data is projected onto the space of solutions of the PDE by the physics loss, which is an effective way of filtering errors \cite{wang_state_2021}.

\subsection{Case 2: Synthetic tomographic images from DNS of channel flow}

\begin{figure}[t]
    \centering
    \includegraphics[width=0.32\textwidth]{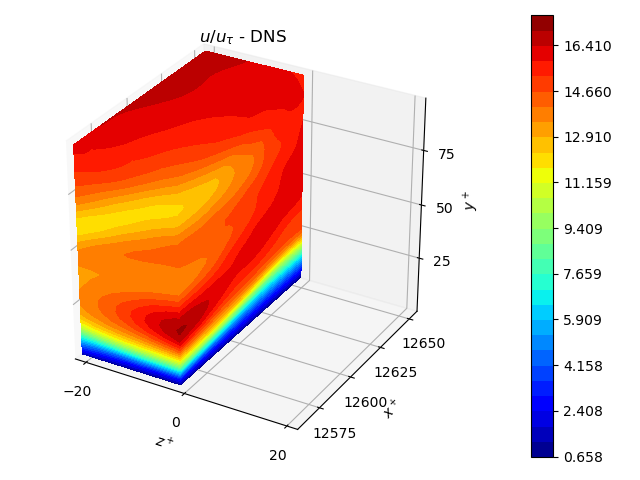}
    \includegraphics[width=0.32\textwidth]{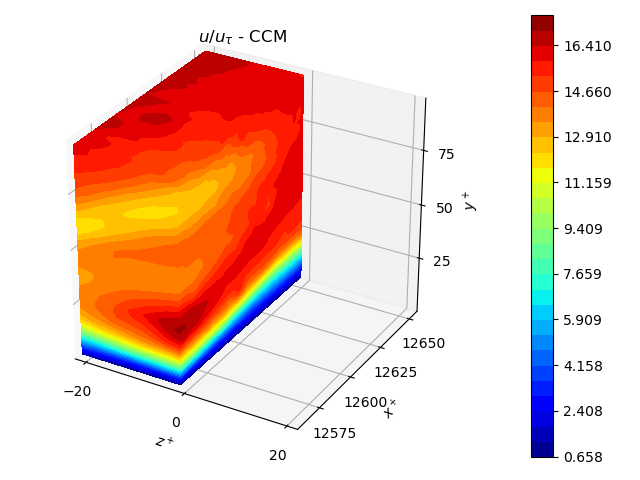}
    \includegraphics[width=0.32\textwidth]{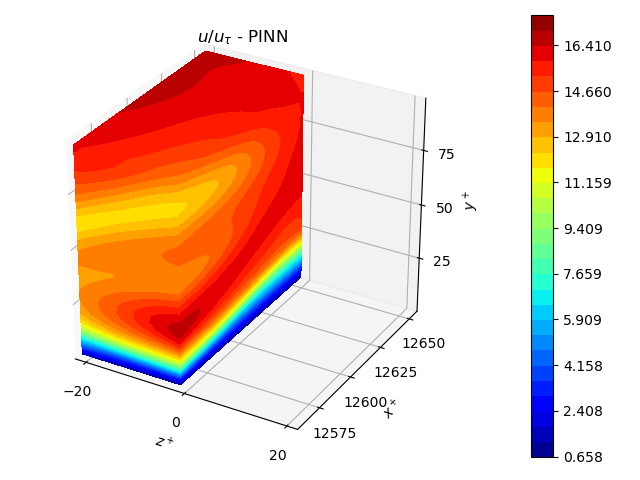}
    \\
    \includegraphics[width=0.32\textwidth]{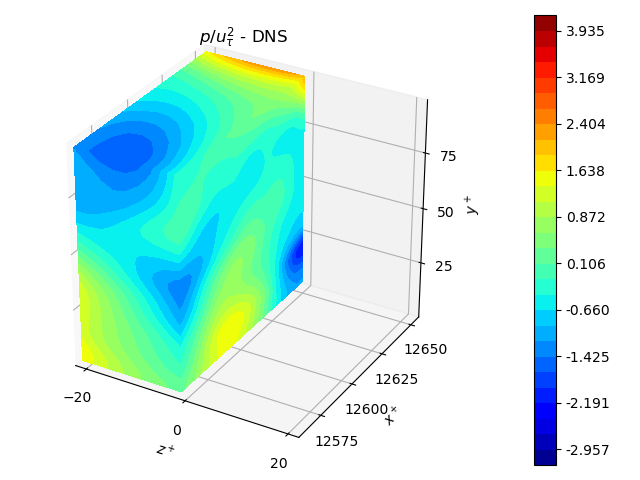}
    \includegraphics[width=0.32\textwidth]{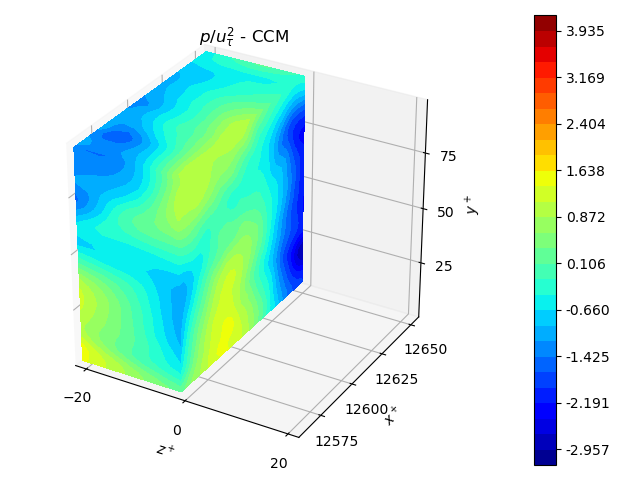}
    \includegraphics[width=0.32\textwidth]{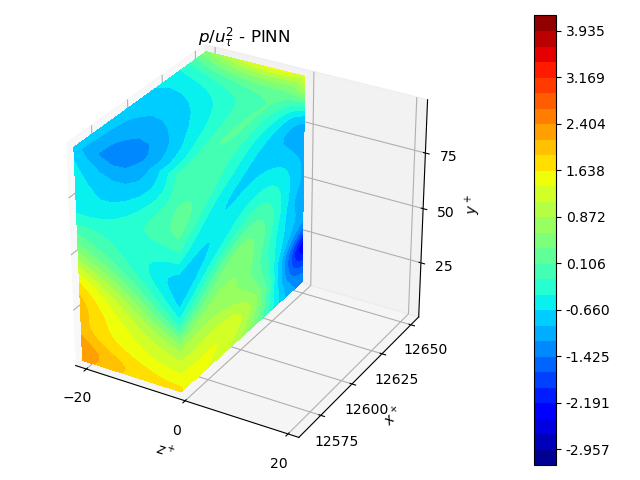}
    \caption{Instantaneous visualization of the data and the CCM and PINN
    reconstruction for Case 2. Top row: streamwise velocity field $u$. Bottom row:
    pressure $p$. Only half of the reconstructed volume is shown.}
    \label{viz_case2}
\end{figure}

\begin{figure}[t]
    \centering
    \includegraphics[width=0.45\textwidth]{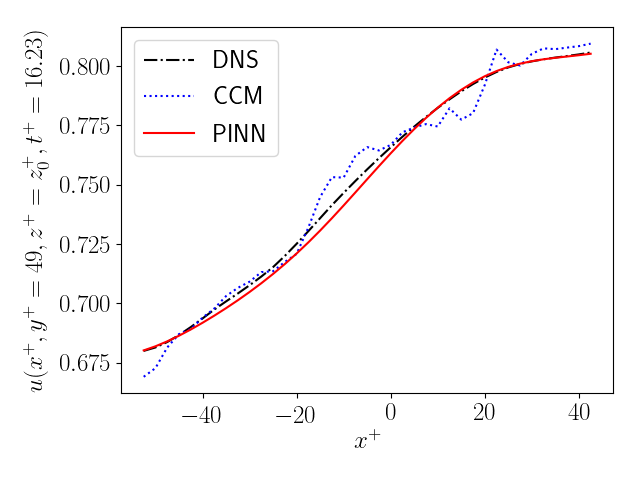}
    \includegraphics[width=0.45\textwidth]{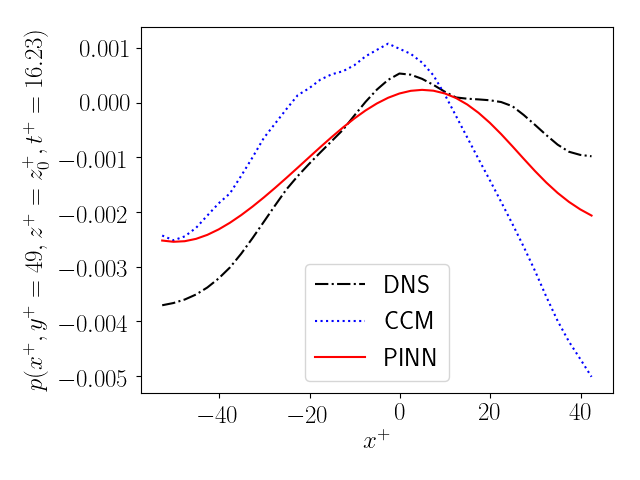}
    \caption{Streamwise profiles extracted from Fig.~\ref{viz_case2}.}
    \label{lines_case2}
\end{figure}

We now turn to Case 2, where we analyze synthetic tomographic images generated using the turbulent channel flow data from the JHTDB. 
Contrary to Case 1 where we performed a reconstruction of an instantaneous field within a very narrow time window (nine snapshots for PINNs and three for CCM), we now reconstruct the time evolution of the flow within the volume. 
In Fig.~\ref{viz_case2} we show visualizations of the streamwise velocity field $u$ and the pressure $p$ of the data, the CCM-reconstructed field and the PINN-reconstructed field at one particular instant. 
The visualizations show only half the volume in order to expose the interior of the field. 
Both CCM and PINNs are in good qualitative agreement with the true data and, similar to the results from Case 1, PINNs produce smoother fields. 
This quality is highlighted in Fig.~\ref{lines_case2}, where we show profiles of instantaneous $u$ and $p$ along the streamwise direction $x$. 
Especially for $u$, the difference in the level of small-scale structure between both techniques is noticeable.

\begin{figure}[t]
    \centering
    \includegraphics[width=0.45\textwidth]{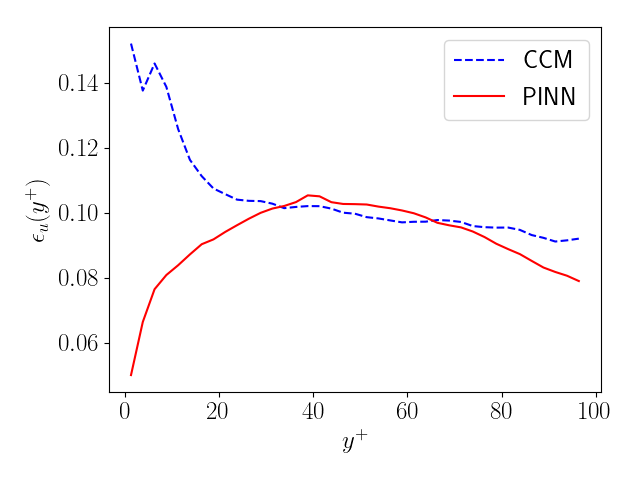}
    \includegraphics[width=0.45\textwidth]{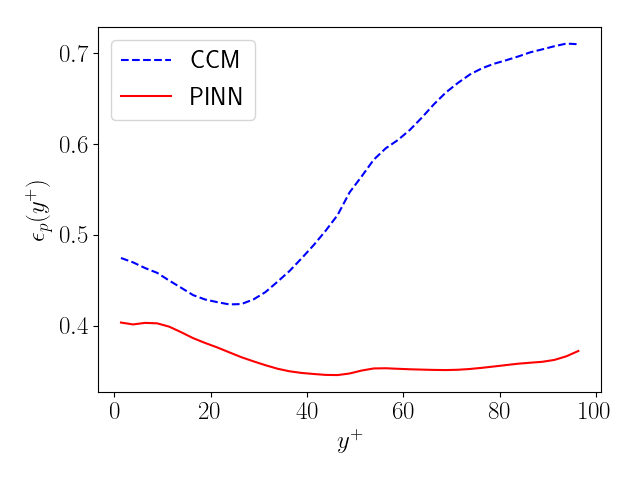}
    \caption{RMSE of streamwise velocity ($\epsilon_u$) and pressure ($\epsilon_p$) as function of height above the wall,   for Case 2.}
    \label{rmse_spatial}
\end{figure}

\begin{figure}[t]
    \centering
    \includegraphics[width=0.45\textwidth]{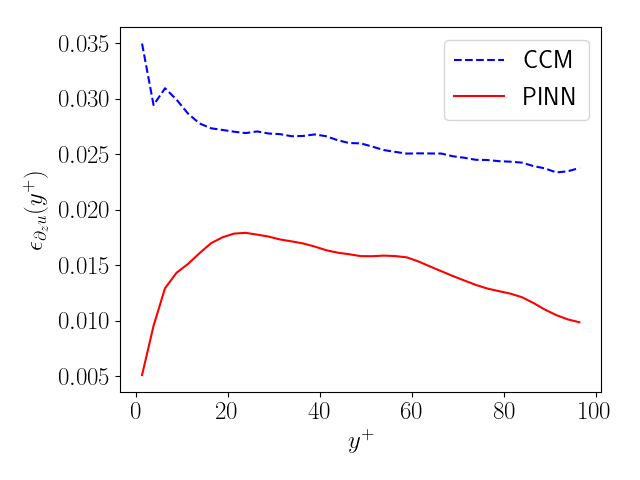}
    \includegraphics[width=0.45\textwidth]{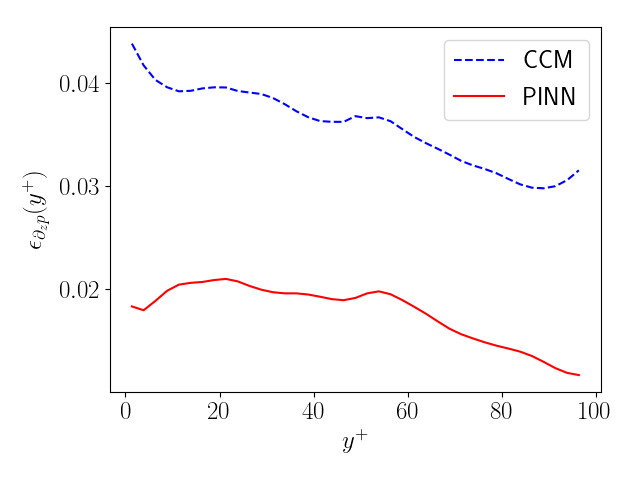}
    \caption{RMSE of spanwise gradient of u-velocity ($\epsilon_{\partial_z u}$) and
    of pressure ($\epsilon_{\partial_z p}$) as function of height above the wall,   for Case 2.}
    \label{rmse_spatial_grads}
\end{figure}

For a quantitative comparison, in Fig.~\ref{rmse_spatial} we show the  RMSEs of the fields as a function of height above the wall, $\epsilon_u(y)$ and $\epsilon_p(y)$, while in Fig.~\ref{rmse_spatial_grads} we show vertical profiles of RMSEs of the spanwise gradients of the fields, $\epsilon_{\partial_z u}$ and $\epsilon_{\partial_z p}$. The error in the reconstruction of $u$ is similar for both techniques, and is on the order of $\SI{0.1}{u_\tau}$, except near the wall where the velocity should vanish. 
PINNs achieve lower errors in the reconstruction of the pressure, especially away from the wall, where the error for the PINNs is approximately half of that of he CCM.
As expected from Fig.~\ref{lines_case2}, the reconstruction errors of the spanwise gradients, of both $u$ and $p$, are significantly lower for the PINNs.

\begin{figure}[t]
    \centering
    \includegraphics[width=0.45\textwidth]{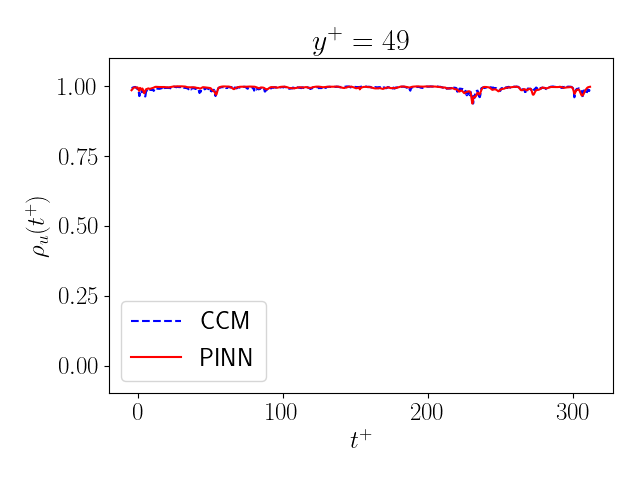}
    \includegraphics[width=0.45\textwidth]{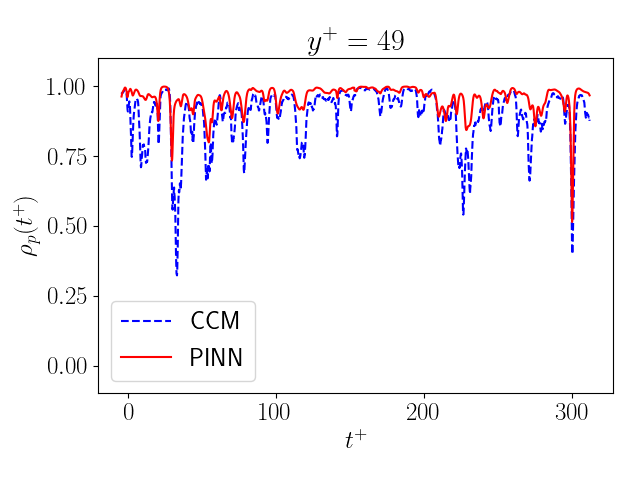}
    \caption{Correlation coefficient between DNS and measured (PINNs and CCM) velocity  and pressure, $\rho_u$ and $\rho_p$, respectively, for Case 2 plotted as function of time.}
    \label{cor_temporal}
\end{figure}

\begin{figure}[t]
    \centering
    \includegraphics[width=0.45\textwidth]{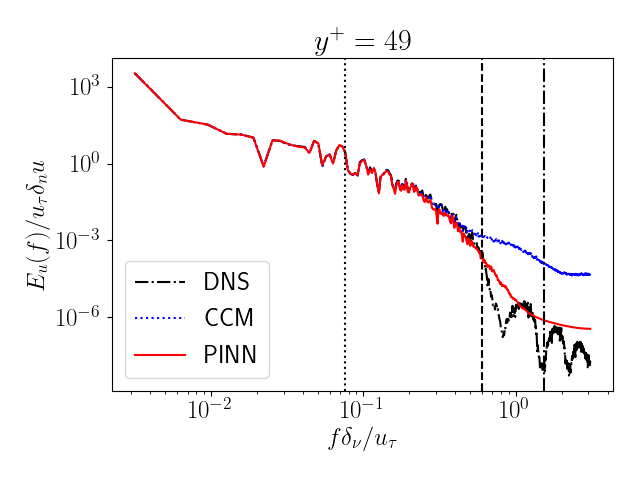}
    \includegraphics[width=0.45\textwidth]{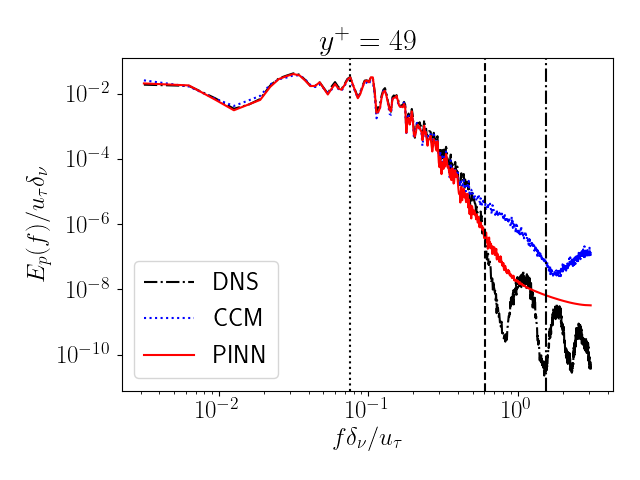}
    \caption{Frequency  spectra of velocity and pressure, $E_u(f)$ and $E_p(f)$, respectively, for Case 2 from DNS and obtained from the PINN and CCM methods.}
    \label{spectra2}
\end{figure}

In order to investigate the temporal behavior of the reconstruction, in Fig.~\ref{cor_temporal} we show the correlation coefficients $\rho_u$ and $\rho_p$ between the true fields and the reconstructed ones, as function of time. 
The reconstructed velocity fields exhibit very high correlations with the true flow throughout the time window, while the reconstructed pressure fields show mostly high accuracy with some instantaneous reductions in the correlations. 
These instantaneous reductions in $\rho_p$ are observed in both CCM and PINNs around the same instants, and the effect is more pronounced in CCM. 
In Fig.~\ref{spectra2} we show the frequency sepctra of the true and reconstructed velocity and pressure, evaluated at $y=49\delta_\nu$.  
A Hanning window was applied to the time signals in order to calculate the spectra, which were then averaged in the horizontal directions. 
Three vertical lines are marked on the figure:  
The dashed line indicates the Nyquist frequency at the height where the spectra are evaluated; this frequnecy is associated with the timescale, $dx/U(y=49\delta_\nu)$ where $dx$ is the streamwise DNS grid size and $U(y=49\delta_\nu)$ is the mean streamwise velocity at this particular height; 
The dotted line indicates the frequency associated with the field of view: $L/U(y=49\delta_\nu)$, where $L$ is the streamwise length of the field of view; 
The dash-dotted line indicates the frequency associated with the temporal resolutions of the JHTDB. 
Both the CCM and PINNs techniques perform very well within the range of timescales of interest. 
The lobes present in the DNS spectra beyond the Nyquist frequency of the time series are due to the temporal interpolation (based on Piecewise Cubic Hermite Interpolating Polynomial) performed by the JHTDB. 
While PINNs are able to smooth out those last frequencies, CCM retains a higher energy content, which is consistent with the observed small-scale fluctuations that can be appreciated in Figs.~\ref{viz_case2} and \ref{lines_case2}.

\begin{figure}[t]
    \centering
    \includegraphics[width=0.32\textwidth]{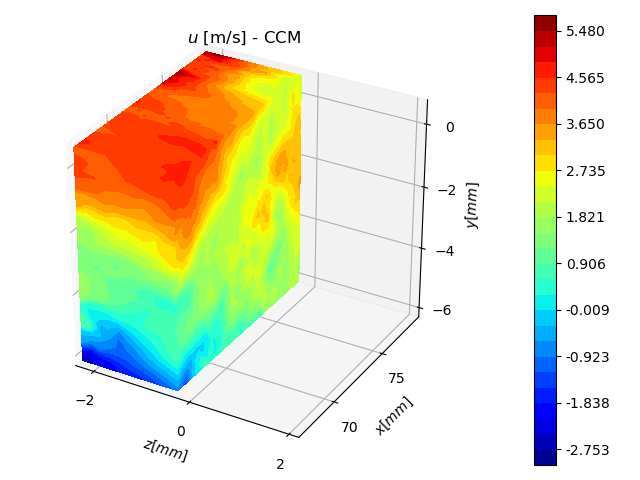}
    \includegraphics[width=0.32\textwidth]{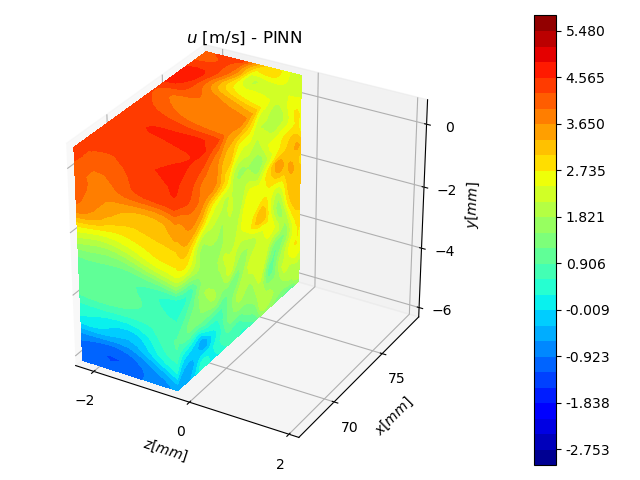}
    \\
    \includegraphics[width=0.32\textwidth]{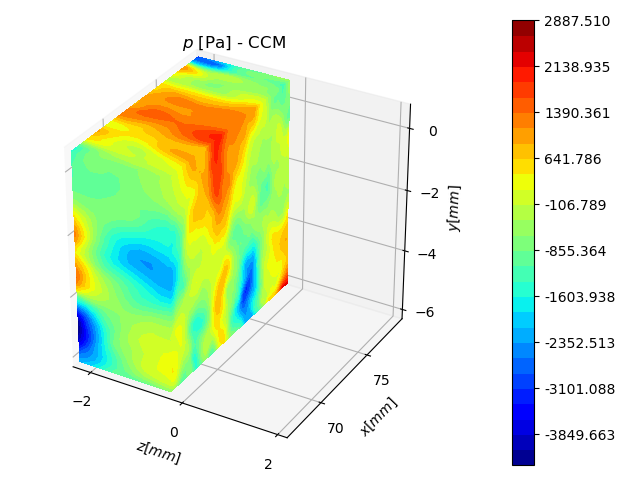}
    \includegraphics[width=0.32\textwidth]{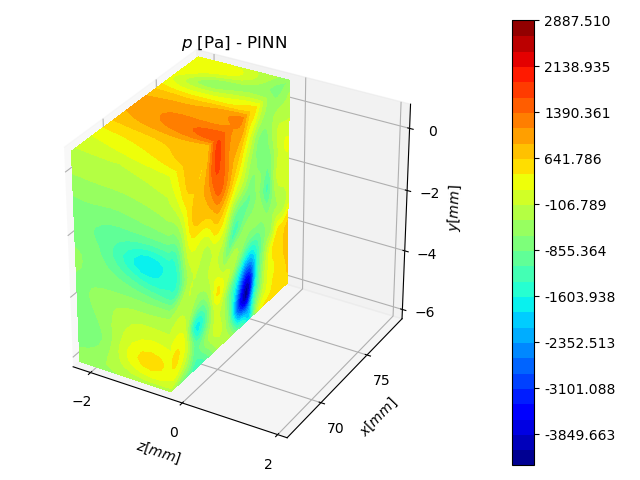}
    \\
    \includegraphics[width=0.32\textwidth]{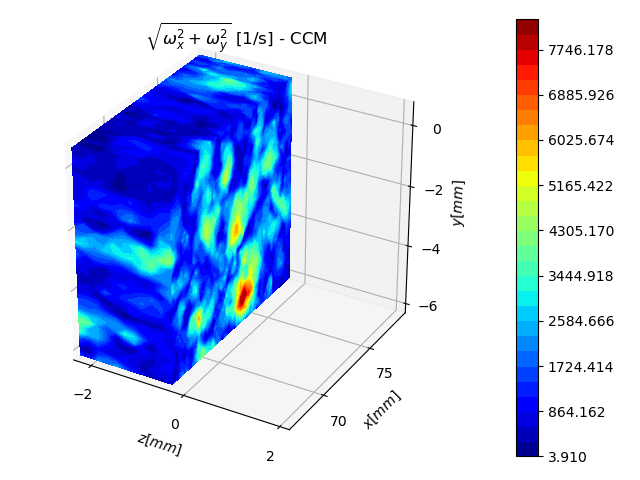}
    \includegraphics[width=0.32\textwidth]{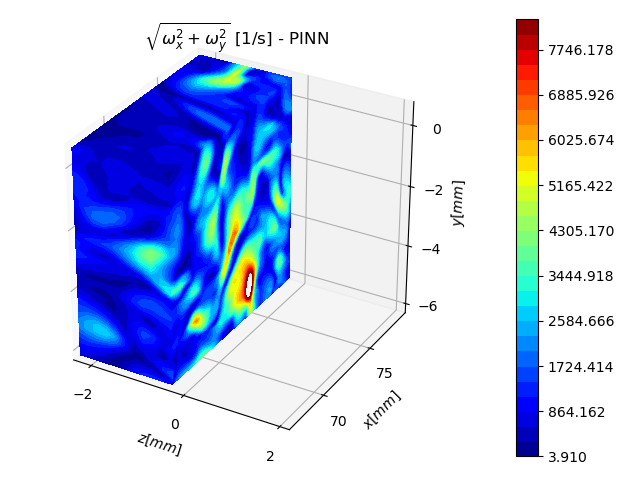}
    \caption{Instantaneous visualization of the CCM and PINN
    reconstruction for Case 3. Top row: streamwise velocity field $u$. Bottom row:
    pressure $p$. Only half of the reconstructed volume is shown.}
    \label{viz_case3}
\end{figure}

\begin{figure}[t]
    \centering
    \includegraphics[width=0.45\textwidth]{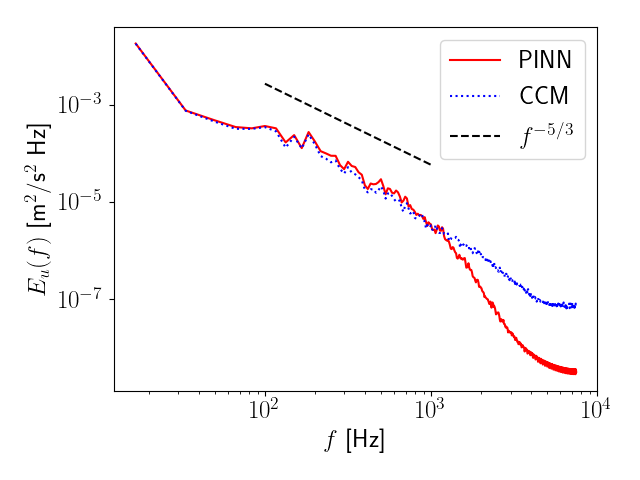}
    \includegraphics[width=0.45\textwidth]{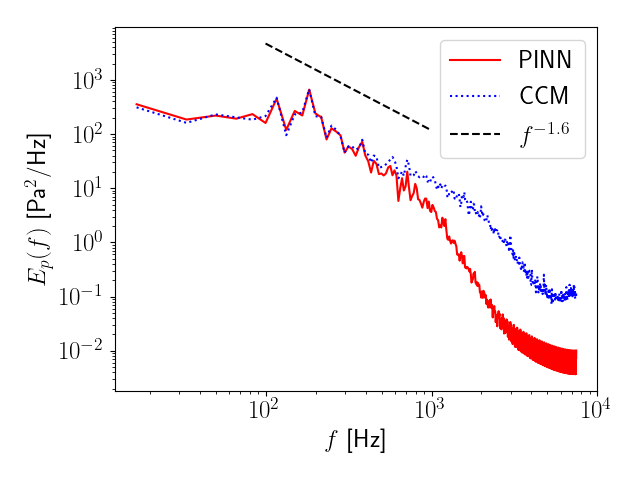}
    \caption{Time-resolved spectra $E_u(f)$ and $E_p(f)$ for Case 3.}
    \label{spectra3}
\end{figure}

\subsection{Case 3: Experimental data in a turbulent shear layer}

In this section we present the results of the reconstruction of the flow behind a step. 
In Fig.~\ref{viz_case3} we show instantaneous visualizations of $u$, $p$ and $\sqrt{\omega_x^2 + \omega_y^2}$ from the CCM and PINN approaches. 
Again only half the volume is visualized in order to expose the interior of the volume. 
Similar to the previous cases, both techniques produce qualitatively similar results, with the locations and shapes of the main structures matching between the two, and with the regions of high vorticity coinciding with the low pressure areas. 
PINNs produce smoother fields as can be expected based on the earlier tests. 
In Fig.~\ref{spectra3} we show the frequency spectra $E_u(f)$ and $E_p(f)$, with reference power-laws $f^{-5/3}$ and $f^{-1.6}$, respectively, as per the literature \cite{tsuji_pressure_2007}. 
The reconstructed spectra match over most frequencies of interest, diverging only at high frequencies as seen in Fig.~\ref{spectra2}. 
The spectra further confirm the trend that PINNs generate smoother results compared to CCM.

\section{Conclusions}

In this paper we present a Physics-Informed Neural Network (PINN) approach for velocity and pressure reconstruction from particle track measurements. The PINN is trained using velocity data extracted from the particle tracks and is regularized using the Navier-Stokes equations. The resulting network learns the specific realization of the flow, i.e., interpolates the velocity field and infers the pressures. We test our approach in three cases, two based on synthetic data from Direct Numerical Simulations (DNS) and one based on experimental measurements, and
compare the PINN results to those obtained by the state-of-the-art Constrained Cost Minimization (CCM) method. PINNs are able to successfully reconstruct the velocity and pressure fields in all cases, achieving errors equal to or smaller than those of CCM. PINNs are also shown to be robust against an increase in both the noise level and sparsity (particle spacing) in the data. Compared to CCM, PINNs produce smoother fields, devoid of small scale noise and thus leading to more reliable predictions of the gradients.

We emphasized a training methodology that can be generalized, characterized by the use of normalization layers and of physical estimates for the weighting parameters to balance the equations. This methodology can be easily applied to other flows or problems, with the proper care and user expertise any reconstruction task requires. The PINN results are robust under small changes in hyperparameters, but an extensive exploration of the effects of changing hyperparameters is left for future work. In Cases 2 and 3 the reconstructed domains were split along the temporal dimension in order to ease training and avoid having to use overly large networks. Such limitations can be mitigated with the use of extended domain PINNs which incorporate domain splitting and training parallelization into their architecture \cite{jagtap_extended_2020}.


\par\bigskip
\noindent
\textbf{Acknowledgements.} 
The authors acknowledge financial support from the Defense Advance Research Projects Agency (AIRA HR00111990025, CompMods HR00112090062), and from the Office of Naval Research (N00014-20-1-2715,  N00014-21-1-2375). The authors thank George Em Karniadakis for useful discussions.

\bibliographystyle{spmpsci}      
\bibliography{ref.bib}   

\end{document}